\documentclass[%
prl,
floatfix,
aps,
superscriptaddress,
nofootinbib,
twocolumn,
showpacs]{revtex4-2}

\usepackage{amsmath,graphicx,amssymb,xcolor,booktabs,braket,multirow,cancel,enumerate,enumitem}
\usepackage[colorlinks=true, allcolors=blue]{hyperref}
\usepackage[normalem]{ulem}
\usepackage{soul}

\usepackage[
    starfontserif 
    ]{starfont}
\usepackage{amsmath}
\allowdisplaybreaks

\DeclareSymbolFont{starfontsym}{OT1}{sts}{m}{n}
\DeclareMathSymbol{\mathSun}{\mathord}{starfontsym}{115}
\DeclareMathSymbol{\mathTerra}{\mathord}{starfontsym}{76}
\DeclareMathSymbol{\mathvarTerra}{\mathord}{starfontsym}{108}

\setstcolor{red}

\newcommand{\red}[1]{{\color{red}#1}}

\newcommand{\orange}[1]{{\color{orange}#1}}

\begin{document}

\title{
Looking for Lights from the Darkness: \\
Signals from MeV-scale Solar Axion-like Particles
}

\author{Yu-Cheng Qiu}
\email{Corresponding author, ethan.qiu@cityu.edu.hk}
\affiliation{Department of Physics, City University of Hong Kong, Kowloon, Hong Kong SAR, China}

\author{Yongchao Zhang}
\email{Corresponding author, zhangyongchao@seu.edu.cn}
\affiliation{School of Physics, Southeast University, Nanjing 211189, China}
\affiliation{Center for High Energy Physics, Peking University, Beijing 100871, China}

\date{\today}

\begin{abstract}
The axion-like particles $a$ can be produced in the Sun via the process of $p + D \to {}^3{\rm He} +a$, with mass up to 5.5 MeV. The photons in the subsequent decay $a \to \gamma\gamma$ can deviate significantly from the Sun, or even from roughly the opposite direction of the Sun. The nontrivial angular and spectral distributions of such photons enable us new methods to detect the {\it lights from the darkness}. In this letter, we consider both the space detection and terrestrial experiments at the South Pole. As a result of the two-body decay and the geometric effects, there exists a critical height for the terrestrial experiments, below which there is no photon for some regions of the parameter space. With the sensitivities of $10^{-16}$ ($10^{-17}$) erg cm$^{-2}$ s$^{-1}$ for the MeV-scale photons in future space and terrestrial experiments, the coupling $g_{a\gamma}$ of $a$ to photons can be probed up to $3\times10^{-12}$ ($1\times10^{-12}$) GeV$^{-1}$, well surpassing the current supernova limits.
\end{abstract}

\maketitle


\noindent {\it Introduction.}---
The quantum chromodynamics (QCD) theory for the strong interaction has an intrinsic CP violating parameter $\bar \theta$, which is tightly constrained by the measured neutron electric dipole moment of $\bar \theta < 10^{-10}$~\cite{Baker:2006ts,Pendlebury:2015lrz,Abel:2020pzs}. The smallness of $\bar \theta$ imposes the 
strong CP problem. One of the well accepted solutions is to promote $\bar \theta$ to be a dynamical field, called the QCD axion. The axion potential is given by the strong dynamics, which stabilizes the field at a CP-even vacuum making $\bar \theta \to 0$ dynamically. 
There are various QCD axion models, whose common residual phenomenology is the anomalous shift symmetry of the axion field, with possible Chern-Simons type coupling with electromagnetism depending on model details~\cite{Peccei:1977hh,Peccei:1977ur,Weinberg:1977ma,Wilczek:1977pj,Kim:1979if,Shifman:1979if,Zhitnitsky:1980tq,Dine:1981rt}.
While the na\"{i}vely heavy QCD axions with mass $m_a \gtrsim \mathcal O (0.1)~{\rm eV}$ are almost excluded~\cite{Raffelt:1990yz,Kim:2008hd,DiLuzio:2020wdo,Carenza:2024ehj,AxionLimits}, 
there are still some phenomenologically viable proposals for MeV-scale or even heavier axions~\cite{Alves:2017avw,Girmohanta:2024nyf,Murayama:2026ioh}.

\begin{figure}
    \centering
    \includegraphics[width=8cm]{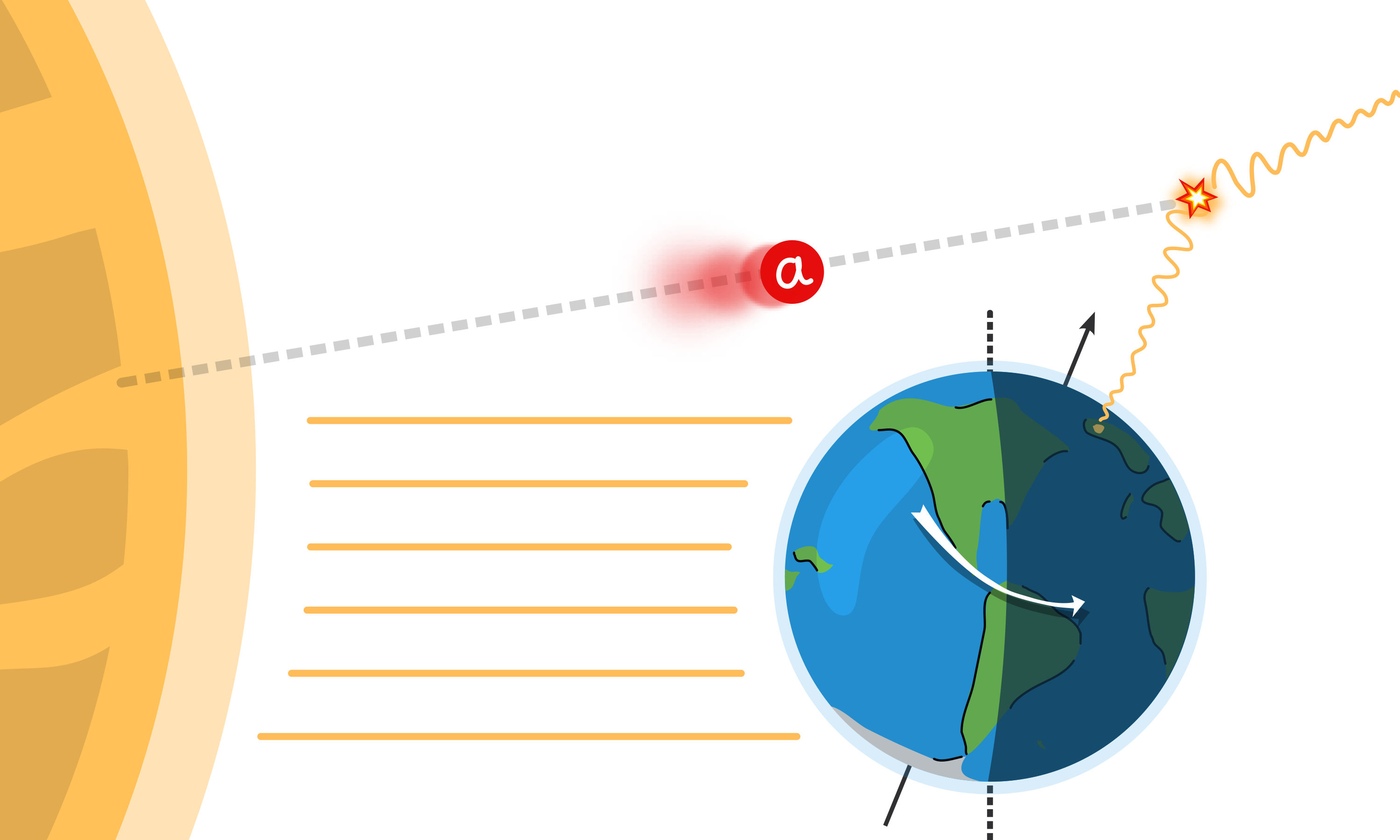}
    \caption{Illustration of the configuration of {\it lights from the darkness},  originating from solar axion decay. The sizes of the Sun and Earth are not to scale. 
    }
    \label{fig:illustration}
    \vspace{-10pt}
\end{figure}

The detection of QCD axion or axion-like particles (ALPs) is essential for the progress of modern fundamental physics~\cite{Adams:2022pbo}.
Here we abuse the term axion for both from now on. 
In addition to the large variety of terrestrial experiments to search for the axions,  
there are also various astrophysical or cosmological data to constrain the axion parameter space~\cite{Graham:2015ouw,Irastorza:2018dyq,Berlin:2024pzi,ParticleDataGroup:2024cfk,AxionLimits}. 
In this letter, 
we propose a new way of looking for the MeV-scale solar axions, via the  the photons from axion decay. 
Due to the two-body decay, there are lights coming from directions that deviated significantly from the Sun, or even from roughly the opposite direction of the Sun. Such a configuration is dubbed here as {\it lights from the darkness}, as illustrated in Fig.~\ref{fig:illustration}. 
The space detectors carried by satellites could pick up a wide angular distribution of the photons from axion decay, which can effectively suppress potential backgrounds directly from the Sun. 
One can also conduct terrestrial experiments in the polar area during the polar night periods.
One distinct feature is the existence of the critical height, below which there is no flux for certain regions of the parameter space.
The expected photon fluxes for both the space and terrestrial detections are estimated, and it turns out that one can explore wide unconstrained regions of the parameter space if the sensitivities of the MeV-scale photon flux can reach the order of $10^{-16}$ erg cm$^{-2}$ s$^{-1}$ in future experiments. 

\noindent {\it MeV-scale solar axion.}---
For phenomenological purpose here,
we consider the following Lagrangian for the axion: 
\begin{align}
    \mathcal L = \frac{1}{2} (\partial_\mu a)^2 - \frac{1}{2} m_a^2 a^2 - \frac{g_{a\gamma}}{4} a F_{\mu\nu} \tilde F^{\mu\nu} - i g_{aN} a \bar N \gamma^5 N\;,
    \label{eq:model}
\end{align}
where $a$ is the axion, $m_a$ is the axion mass, $F_{\mu\nu}$ is the tensor of the electromagnetic field with $\tilde F^{\mu\nu}$ being its dual, and $N = p,\, n$ is the nucleon. For simplicity, we assume that the coupling $g_{aN}$ does not depend on the isospin of nucleons,
and neglect the axion quality problem~\cite{Georgi:1981pu,Giddings:1988cx,Kamionkowski:1992mf,Barr:1992qq,Holman:1992us} and other interactions of axion~\cite{Raffelt:1990yz,Kim:2008hd,DiLuzio:2020wdo,Carenza:2024ehj,AxionLimits}.  

There are multiple channels to produce axions with different masses in the Sun, such as the Primakoff process~\cite{Primakoff:1951iae,Dicus:1978fp} and nuclear transitions~\cite{Donnelly:1978ty,Massarczyk:2021dje,DiLuzio:2021qct}. 
Here we consider the axion with mass $\sim \mathcal O({\rm MeV})$ produced from the $pp$-chain as the benchmark~\cite{Raffelt:1982dr,Gustafson:2023hvm}. 
The main process producing the high energy $\gamma$-rays in the Sun is through a proton $p$ fusing with a deuterium $D$ into a Helium ${}^3{\rm He}$ and releasing a $5.5$~MeV  photon $\gamma$, i.e. $p + D \to {}^3{\rm He} + \gamma$. Given the coupling $g_{aN}$, one can replace the photon $\gamma$ with an axion $a$ if 
$m_a<5.5$~MeV, i.e. $p + D \to {}^3{\rm He} + a$.
The axion flux in this channel can be estimated by $\Phi_a \simeq 0.54 g_{aN}^2 \beta_a^3 \Phi_{\nu pp}$, where $\Phi_\nu pp = 6.0 \times 10^{10}\,{\rm cm}^{-2}\,{\rm s}^{-1}$ is the solar $pp$ neutrino flux, and $\beta_a = \sqrt{1- m_a^2/\omega_a^2}$~\cite{Borexino:2012guz,Raffelt:1982dr}.
The corresponding total axion production rate is
\begin{equation}
\label{eqn:Q}
    \mathcal Q  \simeq 9.1 \times 10^{17} \left(\frac{g_{aN}}{10^{-10}}\right)^2 \beta_a^3 \; {\rm s}^{-1} \,.
\end{equation}

After being produced, the axions can decay into two photons, $a \to \gamma \gamma$, through the interaction in Eq.~\eqref{eq:model}, and the corresponding partial decay width is
\begin{equation}
    \Gamma_{a\to \gamma\gamma}  = \frac{g_{a\gamma\gamma}^2 m_a^4}{64 \pi \omega_a} \equiv {\rm BR}(a\to \gamma\gamma) \, \Gamma_a\;,
\end{equation}
where $\Gamma_a$ is the boosted total decay width.
Here we include the branching ratio (BR) for completeness, and simply take ${\rm BR}(a\to \gamma\gamma) = 1$ for reference throughout this letter. It will change if other decay channels of axion is taken into account, depending on the ultraviolet completion~\cite{Liu:2023bby}.
The decay length of axion is $\ell_a = \beta_a/\Gamma_a$.

\noindent {\it Photon flux.}---
To calculate the (differential) photon flux, we choose the spherical coordinate system $(r,\theta,\varphi)$, with the Sun locating at the center. The detector is located at the position $\vec r_{\rm d} = r_{\rm d} \hat z$ on the $z$-axis, with $r_{\rm d}$ being the distance of the detector from the Sun. The axion decay could happen any time while the axion is traveling outwards.
For simplicity, we assume the axions are produced in a spherical symmetric manner. Then the number density of decaying axions during a time interval ${\rm d}t$ is given by, as a function of $r$, 
\begin{equation}
    {\rm d} n(r) = \frac{\mathcal Q}{4\pi r^2} \frac{e^{-r/\ell_a}}{\ell_a} \, {\rm BR}(a\to \gamma\gamma) \, {\rm d} t\;.
\end{equation}

\begin{figure}
    \centering
    \includegraphics[width=8cm]{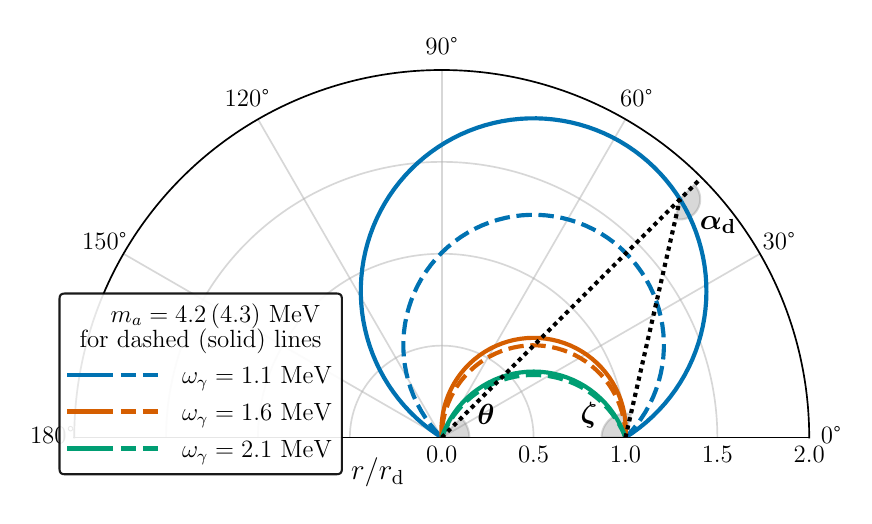}
    \caption{
    Contours of the locations for the two-body decay $a \to \gamma\gamma$ in the $(r,\theta)$ plane, with one of the photons collected by the detector at $r/r_{\rm d} = 1.0$. The dashed and solid contours are for  $m_a = 4.2$ MeV and $4.3$ MeV, and the photon energies of $1.1$ MeV, $1.6$ MeV and $2.1$ MeV are depicted in blue, orange and green, respectively.  
    In the figure $\alpha_{\rm d}$ is the deflection angle, while $\zeta$ is the observation angle.
    }
    \label{fig:geometry}
    \vspace{-10pt}
\end{figure}

For each decaying event, only photons with specific angle can travel to the detector and be collected. The location $(r,\theta,\varphi)$ of the decaying event must satisfy
\begin{equation}
    \frac{r}{r_{\rm d}} = \cos \theta - \frac{\sin\theta}{\tan \alpha_{\rm d}}\;, \label{eq:r_theta}
\end{equation}
where $\alpha_{\rm d}$ is the deflection angle, which is fixed for a given photon energy through the two-body decay kinematics, $2 \omega_\gamma = \omega_a (1-\beta_a^2)/(1-\beta_a \cos \alpha_{\rm d})$. 
The relation in Eq.~(\ref{eq:r_theta}) is axial symmetric, not depending on the azimuthal angle $\varphi$. Therefore, the corresponding axion decaying locations form a curve in the $(r,\theta)$ plane for fixed axion parameters, as exemplified in Fig.~\ref{fig:geometry}. For illustration purposes, we have taken the axion mass to be $4.2$ MeV (dashed) and $4.3$ MeV (solid), and the photon energy to be $\omega_\gamma = 1.1$ MeV (blue), $1.6$ MeV (orange) and $2.1$ MeV (green). 
The observation angle $\zeta$ in Fig.~\ref{fig:geometry} is defined as the angle between the incoming photon momentum and  $\vec r_{\rm d}$.  Then the decaying locations in the $(r,\theta)$ plane can also be expressed as functions of the photon energy $\omega_\gamma$ and the observation angle $\zeta$,
\begin{equation}
    r =  \frac{r_{\rm d}\sin \zeta}{\sin[\alpha_{\rm d}(\omega_\gamma)]} \;,\quad 
    \theta  = \alpha_{\rm d}(\omega_\gamma) - \zeta\,,
    \label{eq:parametrization}
\end{equation}
where $\alpha_{\rm d}$ is a function of $\omega_\gamma$ for fixed axion mass and energy.
The {\it lights from the darkness} configuration corresponds to the axion decay phase space with a large observation angle $\zeta$. As seen in Fig.~\ref{fig:geometry}, such a configuration can be easily obtained, at least in some regions of the parameter space.

For the photons collected by the detector, there is a suppression factor of order $S_{\rm d}/R_{\rm d}^2$, where $S_{\rm d}$ is the effective area of the detector, and $R_{\rm d}$ is the distance between the photon emission location (i.e. the axion decaying location) and the detector, given by $R_{\rm d}^2 = r_{\rm d}^2 + r^2 - 2 r r_{\rm d} \cos \theta$. 
Technically, this suppression factor is obtained by integrating over the solid angle covered by the detector with respect to the axion decaying location. 
The resulting {\it effective} two-body phase space factor for the $a\to \gamma\gamma$ events collected by the detector 
is
\begin{equation}
    \int {\rm d}\, {\rm \Pi}_{{\rm eff}}  \simeq  \frac{1}{4(2\pi)^2} \frac{\omega_\gamma}{\omega_a - \omega_\gamma} \frac{S_{\rm d}}{R_{\rm d}^2} \;.
    \label{eq:detecting_suppression}
\end{equation}
Here we have taken the limit of $S_{\rm d} \ll R_{\rm d}^2$.
In the limit of $R_{\rm d} \to 0$, the axions decay near the detector, 
Eq.~\eqref{eq:detecting_suppression} seems to diverge, and the approximation above breaks down.
However, this divergence can be canceled by the smallness of the phase space of the decaying location near the detector, 
as explained in the Supplemental Material.


The energy of photons collected by the detector is given by the spatial integration of axion decaying events, taking into account the effective phase space in Eq.~\eqref{eq:detecting_suppression}: 
\begin{equation}
    \frac{{\rm d} E_\gamma}{{\rm d}t}  
    = \int_{\mathcal M} \frac{{\rm d}n}{{\rm d}t} r^2 {\rm d}r  {\rm d}\cos \theta {\rm d}\varphi \int {\rm d} \, \Pi_{{\rm eff}} \, \omega_\gamma \;.
\end{equation}
The integration over ${\rm d}r {\rm d}\cos\theta$ can be converted into that over ${\rm d}\omega_\gamma {\rm d}\zeta$, which 
are the physical observables.
The corresponding transformation is
\begin{equation}
    {\rm d}r {\rm d}\cos \theta = {\rm d}\omega_\gamma {\rm d}\zeta \, \left|\frac{\partial \alpha_{\rm d}}{\partial \omega_\gamma} \right| \frac{r_{\rm d}  \sin^2(\alpha_{\rm d} -\zeta )}{\sin^2\alpha_{\rm d}} \;.
    \label{eq:jacobian}
\end{equation}
Denoting $\dot F_\gamma = {\rm d}^2 E_\gamma/{\rm d}S_{\rm d} {\rm d}t$ as the photon energy flux, the corresponding differential rate can be written as
\begin{align}
     {\cal R}_\gamma
     \equiv \frac{{\rm d}^2 \dot F_\gamma}{{\rm d}\omega_\gamma {\rm d}\zeta} & = \frac{\mathcal Q \, {\rm BR}(a\to \gamma\gamma)}{32\pi^2 r_{\rm d} \ell_a} \, f\left(\frac{\omega_\gamma}{\omega_a}\right) e^{-\frac{r(\omega_\gamma,\zeta)}{\ell_a}} \;,
     \label{eq:flux_rate}
\end{align}
where the azimuthal angle $\varphi$ has been integrated out, 
and $r(\omega_\gamma,\zeta)$ 
is given by Eq.~\eqref{eq:parametrization}. 
The dimensionless factor is
\begin{equation}
    f(x) \equiv \frac{x}{1-x} \sqrt{\frac{1-\beta_a^2}{4x(1-x) - (1-\beta_a^2)}}\;,
\end{equation}
and the physical ranges for $\omega_\gamma$ and $\zeta$ are, respectively,
\begin{subequations}
\begin{align}
    \omega_\gamma & \in \left[\frac{\omega_a (1 - \beta_a)}{2}, \, \frac{\omega_a (1 + \beta_a)}{2}\right]\;, \\
    \cos\zeta &\in \left[ \frac{1}{\beta_a} - \frac{\omega_a(1-\beta_a^2)}{2\beta_a \omega_\gamma} , 1  \right]\;.
\end{align}
\end{subequations}
For a given photon energy $\omega_\gamma$, the possible observation angle $\zeta$ 
is bounded from above.  
The differential photon energy rates ${\cal R}_\gamma$ in the $\omega_\gamma$-$\zeta$ plane are presented in Fig.~\ref{fig:flux_contour}, for the benchmark values of {$m_a = 2$ MeV (green), 3 MeV (magenta), 4 MeV (orange) and 5 MeV (blue). The most stringent constraints on MeV-scale axions are from supernova observations, which exclude the coupling $g_{aN} \gtrsim (10^{-11}$--$10^{-10}$) GeV, depending on model details~\cite{Lee:2018lcj,Lella:2023bfb,Lella:2024dmx}. For illustration purpose, we have taken $g_{aN} = 10^{-10}$. The coupling of axion to photons is taken to be $g_{a\gamma} = 5\times10^{-12}$ GeV$^{-1}$, which satisfy the supernova limits~\cite{Payez:2014xsa,Hoof:2022xbe,Muller:2023vjm,Muller:2023pip} {(see Refs.~\cite{Jaeckel:2017tud,Caputo:2022mah,Diamond:2023scc,Diamond:2023cto,Dev:2023hax,Fiorillo:2025yzf,Candon:2025ypl} for weaker limits). 


Here are some important comments, based on Fig.~\ref{fig:flux_contour}. ({\it i}): The flux rate is almost a step function in both $\omega_\gamma$ and $\zeta$, once 
other physical parameters
are fixed. 
Therefore, the averaged differential rate  $\langle {\cal R}_\gamma \rangle$ over the $\omega_\gamma$-$\zeta$ regions with nonzero flux 
is very informative. 
For the parameter space of interest in this letter, the decay length $\ell_a$ is orders of magnitude larger than the distance of Earth from the Sun, for instance
\begin{equation}
\label{eqn:lifetime}
\ell_a \sim 10^5 \, {\rm AU} \, \left( \frac{m_a}{3\,{\rm MeV}} \right)^{-4} \left( \frac{g_{a\gamma}}{10^{-11}\,{\rm GeV}^{-1}} \right)^{2} \,,
\end{equation}
with AU referring to the astronomical unit (AU). 
As a result, the rate $\langle {\cal R}_\gamma \rangle$ becomes orders of magnitude higher when $m_a$ is larger. 
({\it ii}): It is very clear that the maximum observation angle $\zeta$ becomes larger for a smaller $\omega_\gamma$, and a larger photon energy range can be detected for a smaller $m_a$. 
({\it iii}): The observation angle $\zeta$ can be easily much larger than the solar angular radius of $0.27^\circ$ measured at the Earth. This indicates that photons can come from somewhere else other than the direction of the Sun, even from roughly the opposite direction of the Sun  
(cf.~Fig.~\ref{fig:illustration}). Such a geometric feature has been investigated extensively for particles produced in supernovae~\cite{Oberauer:1993yr,Jaffe:1995sw,Jaeckel:2017tud,Mastrototaro:2019vug,Caputo:2021rux,Akita:2022etk,Hoof:2022xbe,Ferreira:2022xlw,Brdar:2023tmi,Akita:2023iwq,Muller:2023vjm,Diamond:2023scc,Syvolap:2024hdh,Telalovic:2024cot,Benabou:2024jlj,Chauhan:2025mnn,Caputo:2025avc,Yu:2026xtb}, neutron stars (NSs)~\cite{Dev:2023hax}, horizontal-branch and Wolf-Rayet stars~\cite{Buckley:2024ldr}, and the starburst galaxy M82~\cite{Candon:2024eah}. 
In light of the huge distances of these stars and galaxy, the corresponding observation angle are expected to be rather small, and the {\it lights from the darkness} configuration is hardly possible~\cite{Hoof:2022xbe}. Only in the vicinity of the Sun, could we have significant angle $\zeta$ as in Fig.~\ref{fig:flux_contour}.  
Here we propose two ways below to detect the photon signals from the darkness, originating from the MeV-scale solar axion decay. 


\begin{figure}
    \centering
    \includegraphics[width=8cm]{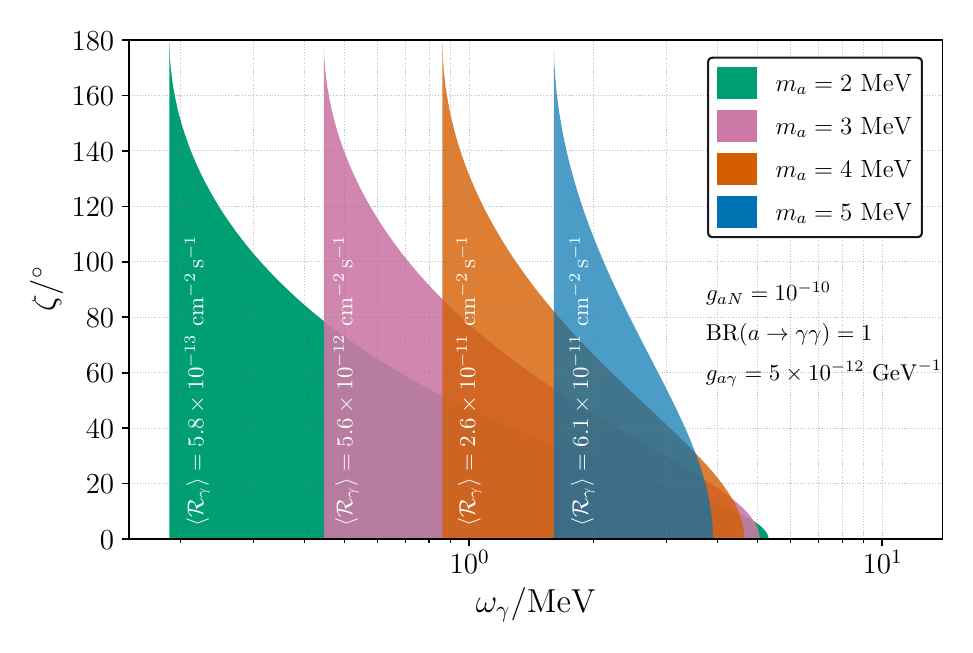}
    \caption{The regions in the $\omega_\gamma$-$\zeta$ plain with non-vanishing photon rate ${\cal R}_\gamma$ in Eq.~(\ref{eq:flux_rate}) for the values of $m_a = 2$ MeV, $3$ MeV, $4$ MeV and $5$ MeV are depicted in green, magenta, orange and blue, respectively.  
    The corresponding averaged differential rates $\langle {\cal R}_\gamma \rangle$ and other benchmark parameters are labeled in the figure. 
    See text for more details.
    }
    \label{fig:flux_contour}
    \vspace{-10pt}
\end{figure}


\noindent{\it The space detection.}---
Suppose we use a space detector loaded on a satellite orbiting around the earth.
With a specific detector design, we can collect the photons from axion decay from almost all over the sky, with a characteristic observation angle distribution as indicated by Fig.~\ref{fig:flux_contour}. One can even receive photons from the direction roughly opposite from the sun for specific parameters.
Taking the same value of $g_{aN} = 10^{-10}$ as in Fig.~\ref{fig:flux_contour}, the expected probable parameter space of  $m_a$ and $g_{a\gamma}$ by the space detectors are shown as the colored regions above the solid and dashed lines in Fig.~\ref{fig:par_space}. 
The darker and lighter regions are for the photon energy ranges of $\omega_\gamma \in [0.01,\,1]$ MeV and $[1,\,5.5]$ MeV, respectively. 
As the axion energy is fixed at $\omega_a = 5.5$ MeV, it is expected that the photons from axion decay tend to have an energy around the MeV scale. Therefore, the photons above 1 MeV are more sensitive to the coupling $g_{a\gamma}$. 
As indicated by the orange and blue lines in Fig.~\ref{fig:par_space}, with the sensitivities of $\dot{F}_\gamma = 10^{-17}$ and $10^{-16}\,{\rm erg}\, {\rm cm}^{-2}\,{\rm s}^{-1}$ for the photon energy range of $[1,\,5.5]$ MeV (two or three orders of magnitude below the designed NuSTAR sensitivities~\cite{NuSTAR:2013yza}), the coupling $g_{a\gamma}$ can be probed up to $1\times10^{-12}$ and $3\times10^{-12}$ GeV$^{-1}$, respectively. 
As implied by Eq.~(\ref{eqn:lifetime}), the photon flux gets larger when axion is heavier, until the factor of $\beta_a^3$ in Eq.~(\ref{eqn:Q}) for axion production  becomes important. As a result, the optimistic sensitivities can be achieved for axion mass $m_a \simeq 4.8$ MeV. 
This can improve the current limits from  SN1987A~\cite{Payez:2014xsa,Hoof:2022xbe,Muller:2023vjm} and SN2023ixf~\cite{Muller:2023pip}, depicted as the gray regions in Fig.~\ref{fig:par_space}, by up to a factor of $\sim 6$.
The photons from solar axion decay have also been searched for by Borexino, but the corresponding limits are very weak~\cite{BOREXINO:2025dbp}.\footnote{The model independent limit from the Borexino data in the channel of $a \to \gamma\gamma$ is $m_a^2 g_{aN} g_{a\gamma} < 1.6 \times 10^{-11}$ eV, orders of magnitude weaker than our prospects in Fig.~\ref{fig:par_space}. The {\it model dependent} limits on $g_{a\gamma}$ in Fig.~7 of Ref.~\cite{BOREXINO:2025dbp} are based on the relation between $g_{aN}$ and $m_a$ in the axion models, and can not be compared directly with our Fig.~\ref{fig:par_space}.}




\begin{figure}
    \centering
    \includegraphics[width=8cm]{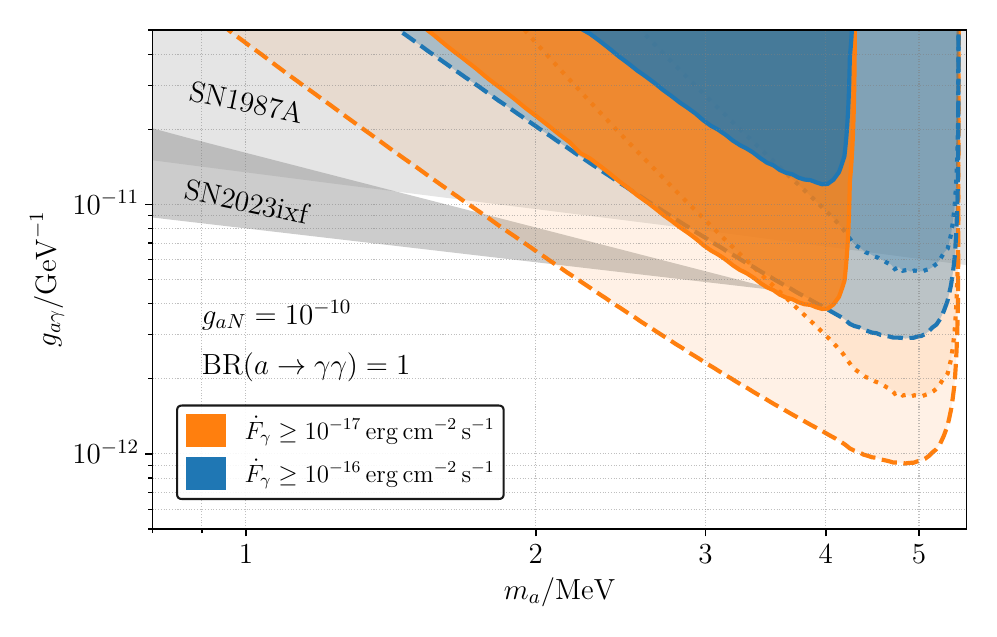}
    \caption{The expected sensitivity regions in the parameter space of $m_a$ and $g_{a\gamma}$ for the searches of photons from solar axion decay.  
    The solid and dashed lines correspond to $\omega_\gamma \in[0.01,1]$~MeV and $[1,5.5]$~MeV, and the orange and blue colors are for the sensitivities of $\dot{F}_\gamma \geq 10^{-17}$ and $10^{-16}\, {\rm erg}\,{\rm cm}^{-2}\,{\rm s}^{-1}$ in the space detections, respectively.
    The dotted lines are the expected sensitivities of the terrestrial experiment at the South Pole at a height of $h = 50$~km.
    The benchmark parameters are labeled in the plot. The gray regions are excluded by the observations of SN1987A~\cite{Payez:2014xsa,Hoof:2022xbe,Muller:2023vjm} and SN2023ixf~\cite{Muller:2023pip}.
    }
    \label{fig:par_space}
    \vspace{-10pt}
\end{figure}

For simplicity, we have neglected all  the potential backgrounds for the MeV-scale photon searches, which may originate directly from the Sun~\cite{Benz:2017abc,Seckel:1991ffa,Fermi-LAT:2011nwz}, the cosmic $X$-ray and $\gamma$-ray backgrounds~\cite{Ueda:2003yx,Hasinger:2005sb,Gilli:2006zi,Fornasa:2015qua}, 
and the interaction of high-energy cosmic rays with atmospheric nuclei~\cite{Gaisser:2016uoy}. Taking into account the two-body decay kinematics, the spectrum and angular distributions of signals are well predicted, as exemplified in Fig.~\ref{fig:flux_contour}, which can help suppress the backgrounds efficiently. However, the detailed analysis are far beyond the main scope of this letter.

\noindent{\it The terrestrial detection.}---
Due to the Earth's axial tilt ($\alpha_\oplus = 23.4^\circ$), the regions with high latitude experience polar nights. To reduce the noise from the Sun, we propose to conduct terrestrial experiments at the Antarctic region during the polar night periods. 
At the ground, the maximum observation solid angle is simply $2\pi$, which is bounded by the tangential plane of the Earth's surface. 
An increase in the observational solid angle can be obtained by lifting the detector to high altitudes using balloons.  
Assuming that the Earth's surface is a perfect sphere and the balloon height is $h$, the observation solid angle is given by $2\pi \left(1+ \sqrt{1- R_\oplus^2 / (R_\oplus+h)^2} \right)$,
where $R_\oplus$ is the radius of Earth. The increased observation solid angle could enhance the photon flux collected by the detector.
Note that here the azimuthal angle $\varphi$ cannot be simply integrated out as in Eq.~\eqref{eq:flux_rate}, since the Earth is blocking part of the photon flux which depends on $\varphi$. The careful treatment is given in the Supplemental Material. 
Taking the Earth's axial tilt into account, the expected fluxes $\dot{F}_\gamma$ above the South Pole at the midnight of its winter solstice 
are presented in Fig.~\ref{fig:pole_flux} as functions of the height $h$. We have chosen the benchmark values of $m_a = 2$ MeV (green), $3$ MeV (magenta), $4$ MeV (orange) and $5$ MeV (blue) and $\omega_\gamma \in [1,\,5.5]$ MeV, and other parameters are the same as in Fig.~\ref{fig:flux_contour}. As expected, the flux can be significantly enhanced when the height gets larger, and eventually reach a plateau when $h \gtrsim R_\oplus$.

\begin{figure}
    \includegraphics[width=8cm]{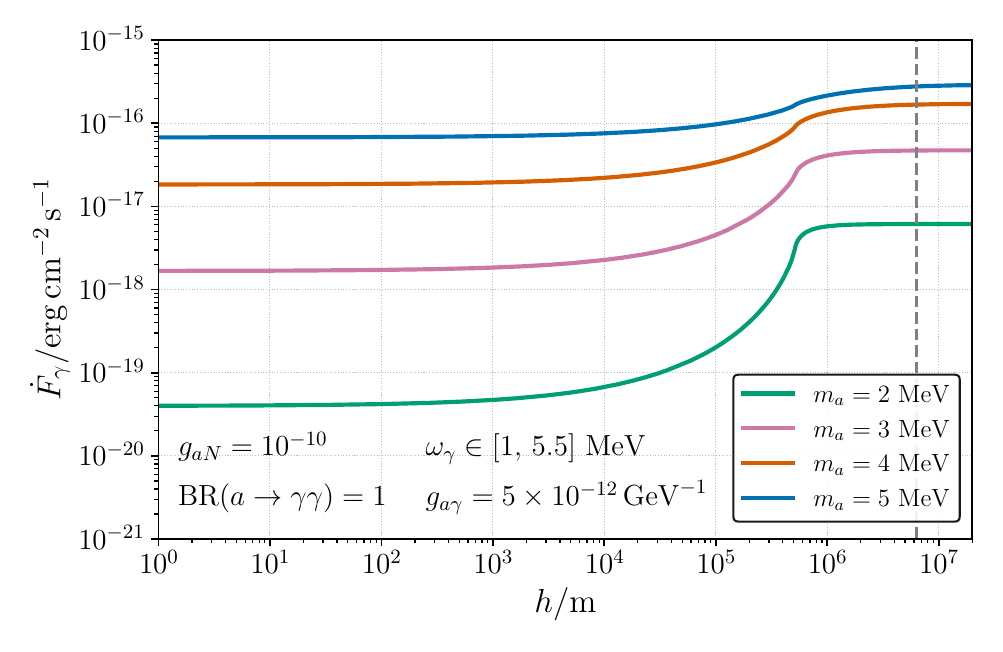}
    \caption{The expected photon energy flux $\dot{F}_\gamma$ as function of the height $h$ at the South Pole at the midnight of its winter solstice. The benchmarks with $m_a = 2$ MeV, 3 MeV, 4 MeV and 5 MeV are depicted in green, magenta, orange and blue, respectively. Other benchmark parameters are labeled in the plot. 
    The vertical gray dashed line indicates the average radius $R_\oplus$ of Earth.
    }
    \label{fig:pole_flux}
    \vspace{-10pt}
\end{figure}

For some specific parameters, the earth itself could block the photon flux totally, depending on the axion mass $m_a$ and photon energy $\omega_\gamma$ (the axion energy $\omega_a$ is fixed at 5.5 MeV). In other words, there exists the critical height $h_{\rm crit}(m_a, \omega_\gamma)$, below which the photon flux vanishes. This can be easily understood qualitatively as follows: when the axion is light, the photons from axion decay tend to be boosted in the forward direction (small deflection angle $\alpha_{\rm d}$), and are less likely to go into the {\it lights from the darkness} configuration.
More details of the critical height can be found in the Supplemental Material. 
As explicit examples, the regions in the $m_a$-$h$ plain with non-vanishing photon fluxes are presented as the shaded regions in Fig.~\ref{fig:critical_height}, for the photon energy ranges of $\omega_\gamma \geq$1.0 MeV (green), 2.4 MeV (magenta), 3.8 MeV (orange) and 4.3 MeV (blue). In this figure, the critical height $h_{\rm crit}$ corresponds to the lower boundaries of the shaded regions. 
For more general cases with experiments not at the South Pole (or the North Pole), the critical hight $h_{\rm crit}$ depends also on the latitudes and seasons of the experimental locations. 
This is a very unique behavior with respect to other astrophysical searches, which can be used to suppress backgrounds maximally. 
Setting $h = 50$ km which is achievable in scientific ballooning~\cite{fuke2017recent}, the expected sensitivity regions of $m_a$ and $g_{a\gamma}$ by the balloon experiments are presented as the dotted lines in Fig.~\ref{fig:par_space} for the photon energy range of $\omega_{\gamma} \in [1,\,5.5]$ MeV,  and the orange and blue colors are for the values of $\dot{F}_\gamma = 10^{-17}$ and $10^{-16}$ erg cm$^{-2}$ s$^{-1}$, respectively. Since the photon signals are partially blocked by the Earth itself, the corresponding sensitivities are to some extent weaker than the space detection. 


\begin{figure}
    \centering
    \includegraphics[width=8cm]{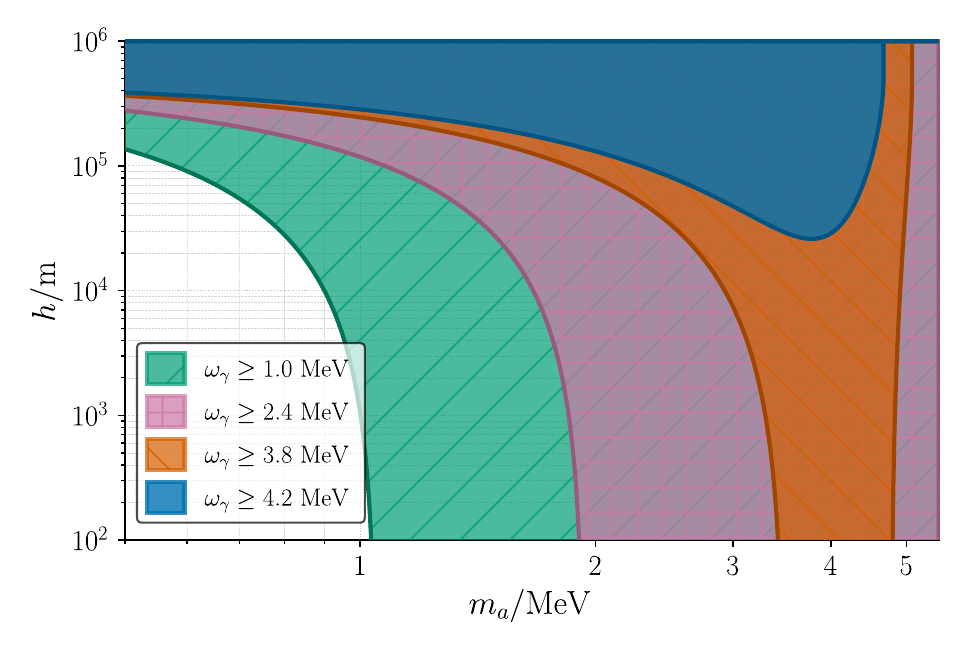}
    \caption{
    The regions of $m_a$ and $h$ with nonzero photon fluxes 
    at the South Pole at the midnight of its winter solstice. The green, magenta, orange and blue regions are for the photon energy ranges of $\omega_\gamma \geq 1.0$ MeV, 2.4 MeV, 3.8 MeV and 4.3 MeV, respectively. 
    The critical height $h_{\rm cirt}$ corresponds to the lower boundaries of the shaded regions, indicated by the darker lines.
    }
    \label{fig:critical_height}
    \vspace{-10pt}
\end{figure}

\noindent{\it Discussions and summary.}---
As a result of Earth's rotation and the revolution around the Sun, the photon signals at the terrestrial experiments from axion decay will have both daily and annual modulation features, which depend on the latitude and height of the experiment. Furthermore, for similar reasons as the critical height $h_{\rm crit}$ in Fig.~\ref{fig:pole_flux}, the signal may vanish within some specific parameter regions. This is dramatically different from the usual modulation signals of dark matter particles~\cite{Freese:2012xd}. More details of the daily and annual modulation signals will be pursued in a future publication.

The nontrivial triangular geometric feature in this letter can be applied to the particles from other compact stars, such as supernovae and NSs, and can also be applied to the decay products of other particles, e.g. the decays of active neutrinos $\nu_j \to \nu_i \gamma$~\cite{Oberauer:1993yr,Jaffe:1995sw}, sterile neutrino $\nu_s \to \nu \gamma, \nu \pi^0,\, \nu \nu \bar\nu,\, \nu \ell^+ \ell^-,\, \ell^\pm \pi^\mp$~\cite{Mastrototaro:2019vug,Brdar:2023tmi,Akita:2023iwq,Chauhan:2025mnn}, CP-even scalar $S \to \gamma\gamma,\, e^+ e^- \gamma$~\cite{Caputo:2021rux,Yu:2026xtb}, Majoron $\phi \to \nu \nu,\, \bar\nu \bar\nu$~\cite{Akita:2022etk,Akita:2023iwq,Telalovic:2024cot}, dark photon $A' \to e^+ e^-,\, \gamma\gamma\gamma$~\cite{Syvolap:2024hdh,Caputo:2025avc} and the gauge boson $Z' \to \nu \bar\nu,\, e^+ e^-$~\cite{Akita:2023iwq}. 
The continuous production of particles at the Sun enables the long-term observations, whereas the particles from core-collapse supernovae and NS mergers are expected to be transient. For the latter case, the time delay as a result of the triangular geometry will provide more information of the decaying mother particles.



In this letter, we focus
on the MeV-scale axions,  
and the energy of photons from axion decay is also of the same order. This encourages the development of the MeV-scale photon detectors for astrophysical purposes, which have been overlooked at the MeV-gap in astronomical observations~\cite{Hunter:2013wla,Wu:2014tya,Wei:2016eox,e-ASTROGAM:2017pxr,Galper:2017cit, VLAST,APT:2021lhj,Orlando:2021get,Caputo:2022xpx,Tomsick:2023aue}. It is expected that the energies of particles from supernovae and NSs can reach higher energy, up to  $\mathcal O(100\,{\rm MeV})$.


In summary, in this letter we have investigated the novel signatures of decaying particles from the Sun as a result of the triangular geometry, taking the MeV-scale axion as a benchmark study. It is found that, in some regions of the phase space, the photons from axion decay could deviate significantly from the original direction of axions, and may even come from roughly the opposite direction of the Sun. The wide angular distribution of the photons is clearly different from the photons directly from the Sun, which can effectively suppress potential backgrounds. 
One can conduct both space and terrestrial experiments to search for the photons from solar axion decay. 
For certain parameter choices, the critical height exists for the terrestrial experiments, below which no secondary photon can be detected. With the sensitivities of $10^{-16}$  ($10^{-17}$) erg cm$^{-2}$ s$^{-1}$ for the MeV-scale $\gamma$-rays, the coupling $g_{a\gamma}$ can be probed up to $3\times10^{-12}$ ($1\times10^{-12}$) GeV$^{-1}$, well surpassing the current supernova limits.





\noindent{\it Acknowledgments.}---
The authors would like to thank Jean-Fran\c{c}ois Fortin,  P. S. Bhupal Dev,  Kuver Sinha, Steven P. Harris, Yuxuan He, Siyang Ling, Ran Li, Yue-Lin Sming Tsai,
Yiming Zhong
for valuable discussions.
The work of Y. -C. Qiu is supported by GRF Grants No.~11302824 and No.~11310925, and CityUHK Grants No.~9610645 and No.~7020130. 
YZ is supported by the National Natural Science Foundation of China under grant No.~12175039 and the ``Fundamental Research Funds for the Central Universities".

\bibliography{ref}
\bibliographystyle{JHEP}


\onecolumngrid
\bigskip
\hrule
\bigskip
\appendix 
\begin{center}
{\bf \large Supplemental Material}
\end{center}

\setcounter{equation}{0}
\setcounter{figure}{0}
\makeatletter
\renewcommand{\theequation}{S\arabic{equation}}
\renewcommand{\thefigure}{S\arabic{figure}}

\section{Flux rate}

The relation of ${\rm d}r {\rm d}\cos\theta$ with ${\rm d}\omega_\gamma {\rm d}\zeta$ in Eq.~\eqref{eq:jacobian} can be written explicitly as
\begin{equation}
    {\rm d}r {\rm d}\cos \theta = {\rm d}\omega_\gamma {\rm d}\zeta \, \frac{\omega_a}{2\omega_\gamma^2} \frac{(1-\beta_a^2) \sin^2(\alpha_{\rm d} - \zeta)}{\beta_a \sin^3 \alpha_{\rm d}}\;, \label{eq:explicit_jacobian}
\end{equation}
where we have used $\beta_a \cos\alpha_{\rm d} = 1 - \omega_a (1-\beta_a^2)/2\omega_\gamma$ from the two-body decay kinematics. Then the differential rate
\begin{align}
    {\cal R}_\gamma & \propto \frac{\omega_a}{2\omega_\gamma^2} \frac{(1-\beta_a^2) \sin^2(\alpha_{\rm d} - \zeta)}{\beta_a \sin^3 \alpha_{\rm d}} e^{-r/\ell_a} 
    \, \frac{\omega_\gamma^2}{\omega_a - \omega_\gamma} \frac{\sin^2\alpha_{\rm d}}{\sin^2(\alpha_{\rm d} -\zeta)}  \int {\rm d} \varphi\;, \nonumber
     \label{eq:flux_rate}
\end{align}
where the constants are neglected here for simplicity, and we have applied the relation $R_{\rm d} = r_{\rm d} \sin(\alpha_{\rm d} - \zeta) / \sin \alpha_{\rm d}$. The divergence in Eq.~\eqref{eq:detecting_suppression} 
corresponds to $\zeta \to \alpha_{\rm d}$ in the denominator, which is canceled by the same factor from the Jacobian in Eq.~\eqref{eq:explicit_jacobian} in the numerator.

\section{Azimuthal angle integration for terrestrial experiments}

For space detection, where we collect the photons from almost all the directions, the integration over the azimuthal angle simply gives $\int {\rm d} \varphi = 2 \pi$. In contrast, for experiments on the ground or near some large celestial bodies, part of the observation solid angle is blocked. Then the integration over the azimuthal angle depends on the observation angle $\zeta$.
For our setup at the South Pole during its winter solstice, the geometric minimal ($\tilde\zeta_{\rm min}$) and maximal ($\tilde\zeta_{\rm max}$) observation angles are functions of the height $h$ but independent of the photon energy, given by
\begin{equation}
    \tilde\zeta_{\rm min}(h)  = \alpha_\oplus - \gamma(h) \;,\quad
    \tilde \zeta_{\rm max}(h)  = \pi + \alpha_\oplus + \gamma(h) \;,
\end{equation}
where $\gamma (h) = \arccos [R_\oplus/ (R_\oplus + h)]$ is the correction from the nonzero height. Here 
$\tilde \zeta_{\rm min}$ can be smaller than zero and $\tilde \zeta_{\rm max}$ can be greater than $\pi$. Intuitively, when the height is very large, i.e. $h \gg R_\oplus$, one has $\gamma \to \pi/2$, which implies that the total range for the observation angle becomes $\tilde \zeta_{\rm max} - \tilde \zeta_{\rm min} \to 2\pi$, as expected. Reversely, on the ground $\gamma \to 0$, and the range for the observation angle is $\pi$. 
For the terrestrial detections not at the South Pole or not on the winter solstice, the values of $\tilde\zeta_{\rm min}$ and $\tilde\zeta_{\rm max}$ will also depend on the experimental locations and the time.

The integration over the azimuthal angle $\varphi$ depends on the observation angle $\zeta$. Let us consider two cases. ({\it i}) the case of $\tilde \zeta_{\rm min} \geq 0$. When $\zeta < \tilde \zeta_{\rm min}$, we could not observe any photons since the Earth itself blocks them, which implies that $\int {\rm d} \varphi = 0$. For $\zeta > 2\pi - \tilde \zeta_{\rm max}$, we can receive the flux from all directions of $\varphi$, which means that $\int {\rm d} \varphi = 2\pi$. In between the two extreme cases, the range for available $\varphi$ increases from $0$ to $2\pi$ as $\zeta$ gets larger. ({\it ii}) the case of $\tilde \zeta_{\rm min}<0$. In this case, there is a boundary $|\tilde \zeta_{\rm min}|$, below which we can collect the flux from all $\varphi$ directions. For $\zeta > 2\pi -\tilde \zeta_{\rm max}$, we have again $\int {\rm d}\varphi = 2\pi$ as above.
The available $\varphi$ range is smaller when $|\tilde \zeta_{\rm min}|\leq \zeta \leq 2\pi - \tilde \zeta_{\rm max}$. Combining these two cases, the most general formula for $\varphi$ integration is given by 
\begin{equation}
    \int {\rm d} \varphi = 
    \begin{cases}
        2\pi H(-\tilde \zeta_{\rm min}(h)) \;, & \zeta < |\tilde \zeta_{\rm min}(h)| \\
        4\arctan \sqrt{\Delta(\zeta,h)} \;, & |\tilde \zeta_{\rm min}(h)| \leq \zeta \leq 2\pi - \tilde \zeta_{\rm max}(h)\\
        2\pi \;, & \zeta > 2\pi - \tilde \zeta_{\rm max}(h)
    \end{cases}\;,
    \label{eq:varphi}
\end{equation}
where $H(x)$ is the Heaviside step function, and
\begin{equation}
    \Delta(\zeta,h) = -
        \frac{\sin [ \zeta - \tilde \zeta_{\rm min}(h) - \gamma(h) ] + \sin [  \gamma(h)]}{ \sin[\zeta + \tilde \zeta_{\rm max}(h)  - \gamma(h)] + \sin [\gamma(h) ]} \;.
\end{equation}
Note that
$\Delta(\zeta,h)$ continuously connects the two limits: in the limit of $\zeta \to \tilde\zeta_{\rm min}>0$ $\Delta$ approaches zero 
due to the vanishing numerator, while $\Delta$ diverges for $\zeta \to - \tilde\zeta_{\rm min}>0$ or $\zeta \to 2\pi - \tilde\zeta_{\rm max}$,
given the vanishing denominator. When $\tilde \zeta_{\rm min} =0$, there will be a singular point at $\zeta = 0$, where $\Delta (\zeta,h)$ is ill-defined. However, this singularity is not physical when integration is performed over ${\rm d} \zeta$, since the limit exists:
\begin{equation}
    \lim_{\zeta \to 0^+} \Delta (\zeta,h; \tilde \zeta_{\rm min } = 0) =  1\;.
\end{equation}

\section{Critical height}

The Earth itself could block some photon flux entirely due to the vanishing of azimuthal angle integration in Eq.~\eqref{eq:varphi} for some ranges of $\zeta$. The two-body decay kinematics tells us that the maximal deflection angle is determined by the smallest photon energy $\omega_\gamma$, which is jointly determined by the axion mass $m_a$ and energy $\omega_a$, apart from the detector capability on the experimental side. 
As implied by Eq.~\eqref{eq:parametrization}, the available maximal observation angle is bounded by the maximal $\alpha_{\rm d}$
for a certain photon energy. If the maximal $\alpha_{\rm d}$ is smaller than a positive $\tilde \zeta_{\rm min}(h)>0$,  
then no flux can be observed, unless $h$ gets larger to obtain a smaller $\tilde \zeta_{\rm min}(h)$ or a larger field of view. 
This defines a critical height $h_{\rm crit}$, which depends only on the axion mass $m_a$ and the lower threshold of $\omega_\gamma$ for fixed $\omega_a$ in our setup of the terrestrial observations at the South Pole. 

The regions of parameter space of $m_a$ and $h$ with vanishing photon flux from axion decay are presented as the gray shaded areas in Fig.~\ref{fig:fluxes}, with the lower thresholds of $\omega_\gamma$ the same as in Fig.~\ref{fig:critical_height}. The expected photon energy flux $\dot{F}_\gamma = 10^{-19}$, $10^{-18}$, $10^{-17}$ and $10^{-16}$ erg cm$^{-2}$ s$^{-1}$ in the $m_a$-$h$ plain are shown as the dotted, dashed, solid and dot-dashed lines in Fig.~\ref{fig:fluxes}, respectively.

\begin{figure}
    \centering
    \includegraphics[width=14 cm]{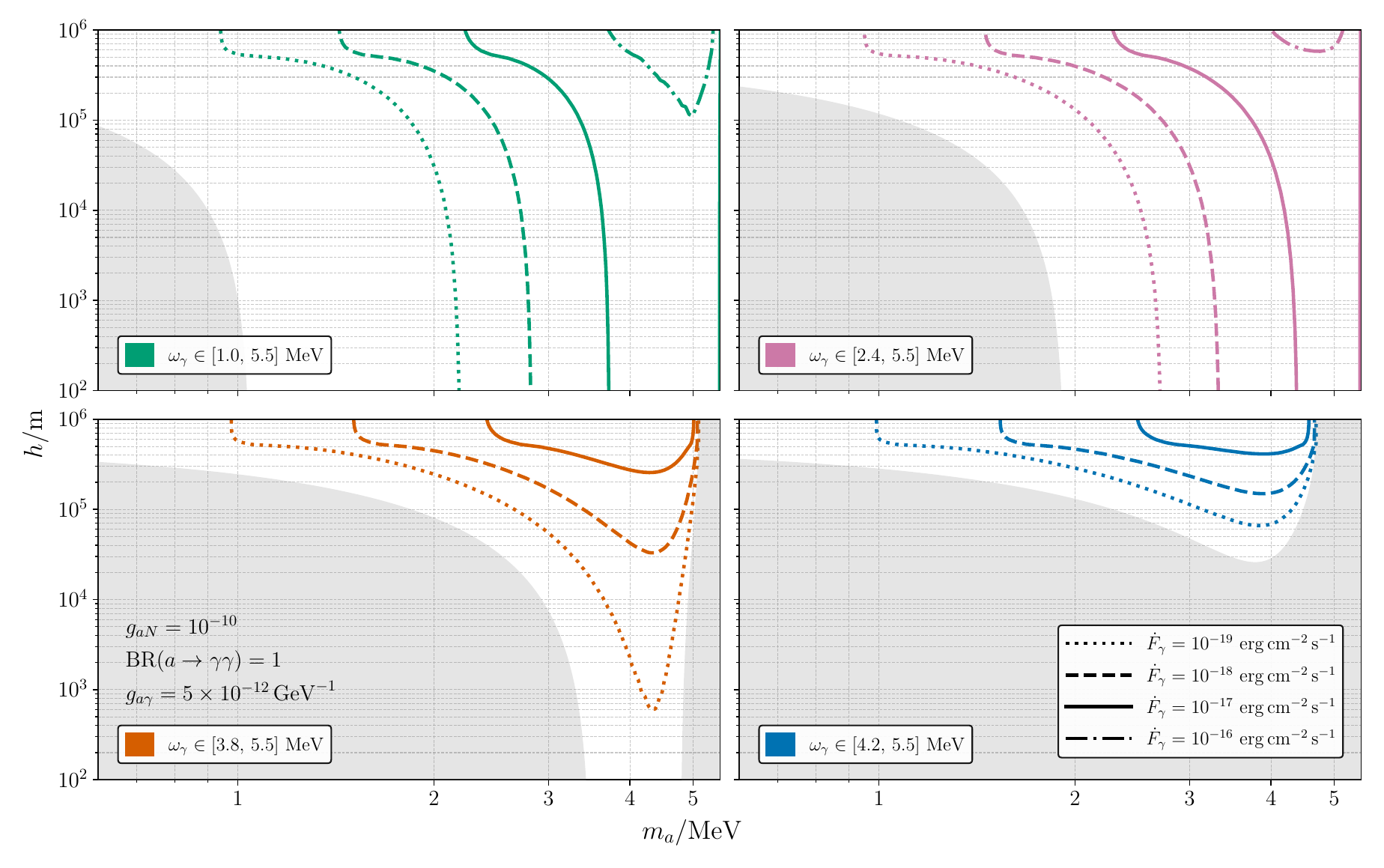}
    \caption{
    Photon energy flux contours $\dot{F}_\gamma = 10^{-19}$, $10^{-18}$, $10^{-17}$ and $10^{-16}$ erg cm$^{-2}$ s$^{-1}$ as functions of $m_a$ and $h$ are depicted as the dotted, dashed, solid and dot-dashed lines respectively. Each panel corresponds to a specific photon energy range as shown in the panels. Other parameters are also shown in the plots.
    The photon flux vanishes in the gray regions (cf. Fig.~\ref{fig:critical_height}). 
    }
    \label{fig:fluxes}
\end{figure}

\end{document}